\documentclass[journal]{IEEEtran}
\usepackage{stfloats}
\usepackage{cite}
\usepackage{float}
\usepackage{graphicx}
\usepackage{subfigure}
\usepackage{color}
\usepackage{algpseudocode}
\usepackage{amssymb}
\usepackage{amsmath}
\usepackage{algorithm}
\usepackage{bm}

\begin{document}

\title{Proactive Monitoring via Jamming in\\
Fluid Antenna Systems}

\author{Junteng Yao, Tuo Wu, Xiazhi Lai, Ming Jin, \emph{Member}, \emph{IEEE}, Cunhua Pan, \emph{Senior Member}, \emph{IEEE},\\
Maged Elkashlan, \emph{Senior Member}, \emph{IEEE}, and Kai-Kit Wong, \emph{Fellow}, \emph{IEEE}

\vspace{-5mm}

\thanks{\emph{(Corresponding authors: Tuo Wu and Cunhua Pan.)}}
\thanks{J. Yao and M. Jin are with the Faculty of Electrical Engineering and Computer Science, Ningbo University, Ningbo 315211, China (e-mail: \{yaojunteng, jinming\}@nbu.edu.cn).}
\thanks{T. Wu and M. Elkashlan are with the School of Electronic Engineering and Computer Science at Queen Mary University of London, London E1 4NS, U.K. (e-mail: \{tuo.wu, maged.elkashlan\}@qmul.ac.uk).}
\thanks{X. Lai is with the School of Computer Science, Guangdong University of Education, Guangzhou, Guangdong, China (E-mail: xzlai@outlook.com).}
\thanks{C. Pan is with the National Mobile Communications Research Laboratory, Southeast University, Nanjing 210096, China. (e-mail: cpan@seu.edu.cn).}
\thanks{K.-K. Wong is with the Department of Electronic and Electrical Engineering, University College London, WC1E 6BT London, U.K., and also with the Yonsei Frontier Laboratory and the School of Integrated Technology, Yonsei University, Seoul 03722, South Korea (e-mail: kat-kit.wong@ucl.ac.uk).}
}

\maketitle

\begin{abstract}
This paper investigates the efficacy of utilizing fluid antenna system (FAS) at a legitimate monitor to oversee suspicious communication. The monitor switches the antenna position to minimize its outage probability for enhancing the monitoring performance. Our objective is to maximize the average monitoring rate, whose expression involves the integral of the first-order Marcum $Q$ function. The optimization problem, as initially posed, is non-convex owing to its objective function. Nevertheless, upon substituting with an upper bound, we provide a theoretical foundation confirming the existence of a unique optimal solution for the modified problem, achievable efficiently by the bisection search method.  Furthermore, we also introduce  a locally closed-form optimal resolution  for maximizing the average monitoring rate. Empirical evaluations confirm that the proposed schemes outperform conventional benchmarks considerably.
\end{abstract}


\section{Introduction}
In the realm of wireless network security, the endeavor to monitor and detect suspicious transmissions has been a focal point of research and development. Xu \emph{et al.} \cite{XuJ16}  made significant strides in this domain by introducing proactive monitoring as a strategy to bolster public security.

This pioneering work laid the foundation for subsequent advancements, including the proposal of proactive monitoring schemes that leverage jamming techniques in relay-assisted networks  \cite{Hudk17}. Recognizing the need for even more efficient monitoring solutions, researchers began to explore reconfigurable intelligent surface (RIS)-assisted proactive monitoring systems \cite{JYao20}. These innovative systems take advantage of the RIS to drastically reduce the power consumption of monitoring  \cite{GHu22}. However, wireless communication is not static in nature and many mission-critical applications are now based on finite blocklength transmissions, posing new challenges for monitoring systems. In response to this shift, investigations have been carried out to design and optimize proactive monitoring systems tailored specifically for finite blocklength transmissions, a topic that has garnered attention in \cite{DXu22,JYao22}. Furthermore, in the context of covert monitoring, Cheng \emph{et al.} \cite{ZCheng21} considered a covert monitoring system, where artificial noise was used.

The aforementioned studies primarily focused on monitors equipped with a single antenna. In pursuit of enhancing monitoring performance, researchers have delved into proactive monitoring schemes with multiple antennas \cite{HCai17,ZhongC17}. Although multi-antenna monitors can indeed offer a substantial improvement in monitoring performance, they are often accompanied by elevated energy consumption and increased hardware costs. These additional expenses primarily stem from the utilization of multiple antennas and radio frequency (RF) chains.

To increase space diversity for performance enhancement while keeping the energy consumption low, a novel technique known as fluid antenna system (FAS) was recently introduced in \cite{KKWong21,KKWong22}. FAS has the ability to flexibly change the antenna position over $N$ preset ports, retrieving space diversity even in a small space. Motivated by this, we propose to incorporate FAS into proactive monitoring. This integration aims to reduce the outage probability of monitor without incurring excessive energy consumption or hardware costs.

In this paper, we aim to maximize the average monitoring rate within the FAS. The primary contributions of this paper can be summarized as follows:
\begin{itemize}
\item We investigate proactive monitoring using FAS in which the legitimate monitor is equipped with a single fluid antenna to monitor suspicious communication. We maximize the average monitoring rate, which involves the integral of the first-order Marcum $Q$ function.
\item The considered problem is non-convex due to the non-convex objective function. Additionally, the expression of this objective function poses significant computational challenges. To address this issue, we employ the upper bound of the objective function, and provide a theoretical proof demonstrating the existence of a unique optimal solution for the formulated problem. This solution can be effectively obtained using a bisection search algorithm.
 \item   To further reduce the computational complexity, we also propose an approximate expression of the objective function, and obtain a closed-form local-optimal solution.
\item Our proposed schemes outperform greatly other benchmarking schemes in terms of monitoring performance, as evidenced by the numerical results.
\end{itemize}

\section{System Model}
Consider a legitimate surveillance FAS configuration comprising the following components: a single-antenna suspicious source (SS), a single-antenna suspicious destination (SD) and a legitimate monitor. The monitor is equipped with a single fluid antenna for receiving signals from the SS and another fixed antenna for transmitting a jamming signal to the SD. The legitimate monitor dynamically switches the fluid antenna to the most favorable port over $N$ evenly distributed ports in a linear space of size $W\lambda$, where $\lambda$ represents the wavelength \cite{KKWong21}. We assume that the monitor knows its own jamming and can flawlessly cancel its self-interference \cite{XuJ16,JYao22}.  Additionally, we consider that any delays resulting from port switching can be negligibly small, e.g., using pixel antennas \cite{KKWong21,KKWong22}.

The channels linking the SS to the SD, the SS to the $k$-th port of the monitor, and the monitor to the SD are denoted as $h$, $g_k$, and $f$, respectively. In this paper, we assume that $h$ and $f$ obey Rayleigh fading distribution, i.e., $h\sim\mathcal{CN}(0,\sigma_h^2)$ and $f\sim\mathcal{CN}(0,\sigma_f^2)$, respectively.

Given the close proximity of the $N$ ports in the FAS, the channels, denoted by $g_k, \forall k$, are inherently correlated. Their mathematical expressions can be defined as \cite{KKWong20,KKWong23}
\begin{align}
g_k = \mu g_0 + (1-\mu)e_k, \quad k = 1,2, \dots, N,
\end{align}
where $N \geq 2$. The channel parameter of a virtual reference port, $g_0$, follows a complex Gaussian distribution denoted by $\mathcal{CN}(0,\sigma_g^2)$ with zero mean and variance $\sigma^2$. The term $e_k$ is an independently and identically distributed (i.i.d.) random variable, also distributed as $\mathcal{CN}(0,\sigma_g^2)$. Additionally, $\mu$ represents the correlation factor, which is given by \cite{KKWong20,KKWong23}
\begin{align}\label{e2}
\mu = \sqrt{2}\sqrt{{}_1F_2\left(\frac{1}{2};1;\frac{3}{2};-\pi^2W^2\right)-\frac{J_1(2\pi W)}{2\pi W}}
\end{align}
with ${}_a F_b$ being the generalized hypergeometric function and $J_1(\cdot)$ as the first-order Bessel function of the first kind.

Given $|g_0|$, the probability density function (PDF) of $g_k$ can be expressed as
\begin{align}\label{e3}
f_{|g_k| \big| |g_0|}(r_k | r_0) = \frac{2r_k}{\sigma_g^2 (1-\mu^2)} e^{-\frac{r_k^2+\mu^2 r_0^2}{\sigma_g^2 (1-\mu^2)}} I_{0}\left(\frac{2\sigma_g \mu r_k r_0}{\sigma_g^2 (1-\mu^2)}\right),
\end{align}
where $I_{0}(u)$ denotes the modified Bessel function of the first kind and order zero. Its series representation is given by \cite{ISGradshteyn07}
\begin{align}\label{eq4}
I_0(z) = \sum_{k=0}^{\infty} \frac{z^{2k}}{2^{2k} k!\Gamma(k+1)}.
\end{align}

The SS broadcasts a signal, denoted as $x_s$ with $\mathbb{E}[|x_s|^2]=1$. The received signals at the SD and the $k$-th port of the legitimate monitor are, respectively, expressed as
\begin{align}
y_d &=\sqrt{p_s}hx_s+\sqrt{p_m}fx_m+n_d,\\
\label{q2}y_{m,k} &=\sqrt{p_s}g_kx_s+n_m,
\end{align}
where $p_s$ represents the transmission power of the SS; $p_m$ denotes the power of the jamming signal; $x_m$ represents the jamming signal with $\mathbb{E}[|x_m|^2]=1$ sent from the monitor, $n_d\sim \mathcal{CN}(0,\sigma_d^2)$ and $n_m\sim\mathcal{CN}(0,\sigma_m^2)$ denote the additive Gaussian noise at the SD and the monitor, respectively.

Therefore, the signal-to-interference-plus-noise ratio (SINR) at the SD and the signal-to-noise ratio (SNR) at the $k$-th port of the monitor  can be expressed as
\begin{align}\label{q4}
\gamma_d&=\frac{p_s|h|^2}{p_m|f|^2+\sigma^2_d},\\
\label{q5}
\gamma_{m,k}&=\frac{p_s|g_k|^2}{\sigma^2_m},
\end{align}
respectively.

We assume that the monitor can promptly  switch the fluid antenna to the most favorable port,  and the maximum of $|g_k|$ can be expressed as
\begin{align}
|g_{\max}| =\max\{|g_1, |g_2|, \dots, |g_N|\}.
\end{align}
As a result, the SNR of the monitor is found as
\begin{align}\label{q5-1}
\gamma_{m}&=\frac{p_s|g_{\max}|^2}{\sigma^2_m}.
\end{align}

In accordance with \cite{XTang16}, for a given transmission rate $R$, the outage probability at the SD is defined as
\begin{align}\label{q6}
\mathbb{P}^{\mathrm{out}}_{d}&=\mathbb{P}(\log_2(1+\gamma_d)<R)=1-\frac{\lambda_2}{\lambda_1+\lambda_2}e^{-\lambda_1\sigma^2_h\gamma_{th}},
\end{align}
where $\gamma_{th}=2^R-1$, $\lambda_1 = \frac{1}{\sigma^2_hp_s}$, and $\lambda_2 = \frac{1}{\sigma^2_f\gamma_{th}p_s}$.

According to \cite[(16)]{KKWong20}, the outage probability at the monitor for a given transmission rate $R$ is expressed as
\begin{align}\label{q7}
\mathbb{P}^{\mathrm{out}}_{m}=&\mathbb{P}(\log_2(1+\gamma_m)<R)\nonumber\\
=&\int_0^\infty e^{-t} \nonumber\\
&\times \left[1-Q_1\left(\sqrt{\frac{2\mu^2}{1-\mu^2}}\sqrt{t},\sqrt{\frac{2}{1-\mu^2}}\sqrt{\frac{\gamma_{th}}{\Gamma}}\right)\right]^{N}dt,
\end{align}
in which $\Gamma=p_s \sigma^2_g/\sigma^2_m$ and $Q_1(\cdot,\cdot)$ represents  the first-order Marcum $Q$-function. The average monitoring rate at the legitimate monitor can therefore be determined by \cite{XuJ16}
\begin{equation}\label{eq13}
R_m = R(1-\mathbb{P}^{\mathrm{out}}_{m}).
\end{equation}

Our objective is to optimize  $p_m$ for maximizing the average monitoring rate at the legitimate monitor. This leads to the following optimization problem:
\begin{align}\label{q11}
\max_{0\leq p_m \leq p^{\max}_m}\ \ R_m\ \ \mbox{s.t.}\ \mathbb{P}^{\mathrm{out}}_{d}=\delta,
\end{align}
where $p^{\max}_m$ represents the maximum transmission power of the jamming signal, and $\delta > 0$ denotes the target outage probability at the SD. Problem \eqref{q11} is a non-convex problem due to the non-convex objective function, which is challenging.

\section{Proactive Monitoring using an Upper Bound}
Here, we begin by deriving an upper bound of the objective function of \eqref{q11}. Subsequently, we prove that this optimization problem possesses a unique solution, which can be effectively determined using the bisection search method.

According to \eqref{eq4}, it can be seen that a lower bound of the $f_{|g_k| | |g_0|}(r_k | r_0)$ can be found by
\begin{align}\label{q7-1}
f_{|g_k| | |g_0|}(r_k | r_0)=&\frac{2r_k}{\sigma_g^2 (1-\mu^2)} e^{-\frac{r_k^2+\mu^2 r_0^2}{\sigma_g^2 (1-\mu^2)}}.
\end{align}

Considering  $|g_0|$, it is evident that $|g_1|, \dots, |g_N|$ are mutually independent. This allows us to deduce the joint PDF of $|g_1|, \dots, |g_N|$ as
\begin{align}\label{eq16}
f_{|g_1|, \dots, |g_N| | |g_0|}(r_1, \dots, r_N| r_0)=&\prod_{k=1}^N\frac{2r_k}{\sigma_g^2 (1-\mu^2)} e^{-\frac{r_k^2+\mu^2 r_0^2}{\sigma_g^2 (1-\mu^2)}}.
\end{align}
Also, from \eqref{eq16}, the joint PDF of $|g_0|, \dots, |g_N|$ is written as
\begin{align}\label{eq16-1}
f_{|g_0|, \dots, |g_N|}(r_0, \dots, r_N)=&\frac{2r_0}{\sigma_g^2} e^{-\frac{r_0^2}{\sigma_g^2}}\prod_{k=1}^N\frac{2r_k}{\sigma_g^2 (1-\mu^2)} e^{-\frac{r_k^2+\mu^2 r_0^2}{\sigma_g^2 (1-\mu^2)}}.
\end{align}
Now, using \eqref{eq16}, the lower bound of $\mathbb{P}^{\mathrm{out}}_{m}$ can be written as
\begin{align}\label{eq17}
\hat{\mathbb{P}}^{\mathrm{out}}_{m}=&\mathbb{P}\left(|g_0|<\infty, |g_1|<\sqrt{\frac{\gamma_{th}\sigma_g^2}{\Gamma}},\dots,|g_N|<\sqrt{\frac{\gamma_{th}\sigma_g^2}{\Gamma}}\right)\nonumber\\
=&\eta \left[1-e^{-\frac{\gamma_{th}}{\Gamma(1-\mu^2)}}\right]^N,
\end{align}
where $\eta = \frac{1-\mu^2}{1+(N-1)\mu^2}$. Therefore, an upper bound of the average monitoring rate  $R_m$ can be found as
\begin{equation}\label{eq18}
\hat{R}_m = R(1-\hat{\mathbb{P}}^{\mathrm{out}}_{m}).
\end{equation}
As a result, Problem  \eqref{q11}  can be reformulated by
\begin{align}\label{eq19}
\max_{0\leq p_m \leq p^{\max}_m}\ \ \hat{R}_m\ \ \mbox{s.t.}\ \mathbb{P}^{\mathrm{out}}_{d}=\delta.
\end{align}

Based on \cite{XuJ16}, it is established that $R$ and $p_m$ are in one-to-one correspondence subject to the constraint $\mathbb{P}^{\mathrm{out}}_{d}=\delta$ in Problem \eqref{eq19}. Consequently, the constraint $0\leq p_m \leq p^{\max}_m$ can be equivalently expressed as
\begin{equation}
R^{\min}\leq R \leq R^{\max}
\end{equation}
with
\begin{align}
R^{\min}&=\log_2\left(1+\frac{p_s\sigma^2_h}{\sigma^2_d}\mathcal{W}\left(\frac{\sigma^2_de^{\frac{\sigma^2_h}{p_m\sigma^2_f}}}{\sigma^2_fp_m(1-\delta)}\right)-\frac{\sigma^2_hp_s}{p_m\sigma^2_f}\right),\\
R^{\max}&=\log_2\left(1-\frac{p_s\sigma^2_h\ln(1-\delta)}{\sigma^2_d}\right),
\end{align}
where $\mathcal{W}(z)$ represents the Lambert $\mathcal{W}$ function.

After obtaining $R^{\min}$ and $R^{\max}$, \eqref{eq19} is recast as
\begin{align}\label{eq23}
\max_{R}\ \ \hat{R}_m\ \ \mbox{s.t.}\ R^{\min}\leq R \leq R^{\max}.
\end{align}
To solve \eqref{eq23}, we let $x=2^R-1$ and have $R=\log_2(1+x)$. The objective function $\hat{R}_m$ can be rewritten as
\begin{align}\label{q27}
F(x)=\log_2(1+x)\left(1-\eta\left(1-e^{-\frac{x}{\Gamma(1-\mu^2)}}\right)^N\right).
\end{align}

Upon calculating the first-order partial derivative of $F(x)$ with respect to $x$, and performing subsequent mathematical manipulations, we obtain the following results:
\begin{align}\label{qq23}
\frac{\partial F(x)}{\partial x}=&h(x)-g(x),
\end{align}
where
\begin{align}
\label{eq26}h(x) =& \frac{1}{\ln2(1+x)}\left(1-\eta\left(1-e^{-\frac{x}{\Gamma(1-\mu^2)}}\right)^N\right),\\
\label{eq27}g(x)=&\frac{N\log_2(1+x)}{\Gamma(1-\mu^2)}\left(\eta\left(1-e^{-\frac{x}{\Gamma(1-\mu^2)}}\right)^{N-1}e^{-\frac{x}{\Gamma(1-\mu^2)}}\right).
\end{align}
To continue, we have the following lemma.

\emph{Lemma 1}: $F(x)$ is a monotonically increasing function over the interval $x\in[0,x^o]$, and monotonically decreasing function over the interval $x\in(x^o, +\infty]$, where $x^o=2^{R^o}-1$ and $R^o$ is the solution of the equation  $\frac{\partial \hat{R}_m}{\partial R}=0$.

\begin{proof}
See Appendix A.
\end{proof}

According to Lemma 1, we can ascertain that Problem \eqref{eq19}  possesses a distinctive optimal solution, which is given by
\begin{align}\label{p2}
\min\{\max\{R^{\min}, R^o\}, R^{\max}\},
\end{align}
where $R^o$  can be determined through utilizing the bisection search method within the interval $R\in[R^{\min},R^{\max}]$.

\emph{Complexity Analysis}: The computational complexity of solving Problem \eqref{eq19} is
\begin{equation}
\mathcal{O}\left(\log_2 \left(\frac{R^{\max}-R^{\min}}{\epsilon}\right)\right),
\end{equation}
where $\epsilon$ denotes the accuracy of bisection search.

\section{A Closed-Form Solution to Proactive Monitoring}
In this section, we aim to reduce the computational complexity for achieving proactive monitoring by proposing a closed-form solution.

Considering $0 \leq Q_1\left(\sqrt{\frac{2\mu^2}{1-\mu^2}}\sqrt{t},\sqrt{\frac{2}{1-\mu^2}}\sqrt{\frac{\gamma_{th}}{\Gamma}}\right)\leq 1$, we can obtain
\begin{multline}\label{eq30}
\left[1-Q_1\left(\sqrt{\frac{2\mu^2}{1-\mu^2}}\sqrt{t}\sqrt{\frac{2}{1-\mu^2}}\sqrt{\frac{\gamma_{th}}{\Gamma}}\right)\right]^{N}\\
\approx 1-NQ_1\left(\sqrt{\frac{2\mu^2}{1-\mu^2}}\sqrt{t},\sqrt{\frac{2}{1-\mu^2}}\sqrt{\frac{\gamma_{th}}{\Gamma}}\right).
\end{multline}
Therefore, $\mathbb{P}^{\mathrm{out}}_{m}$ can be approximated by
\begin{align}\label{eq31}
\tilde{\mathbb{P}}^{\mathrm{out}}_{m}=&1-Ne^{-\frac{\gamma_{th}}{\Gamma}}.
\end{align}

The approximated average monitoring rate at the legitimate monitor is denoted as
\begin{equation}\label{eq32}
\tilde{R}_m = R(1-\tilde{\mathbb{P}}^{\mathrm{out}}_{m})=NRe^{-\frac{\gamma_{th}}{\Gamma}}.
\end{equation}
This expression \eqref{eq32} reveals that the approximated average monitoring rate is essentially $N$ times $Re^{-\frac{\gamma_{th}}{\Gamma}}$. This is consistent with the average monitoring rate of the conventional single antenna scheme presented in \cite{XuJ16}.

The optimization  problem is thus expressed as
\begin{align}\label{eq33}
\max_{0\leq p_m \leq p^{\max}_m}\ \ Re^{-\frac{\gamma_{th}}{\Gamma}}\ \ \mbox{s.t.}\ \mathbb{P}^{\mathrm{out}}_{d}=\delta.
\end{align}

Analogous to \emph{Lemma 1}, we find that $Re^{-\frac{\gamma_{th}}{\Gamma}}$  first increases and then decreases concerning $R$. The closed-form solution for $\frac{\partial \tilde{R}_m}{\partial R}=0$ is therefore provided by
\begin{align}\label{eq34}
\bar{R}^\star=\frac{1}{\ln2}\mathcal{W}\left(\Gamma \right).
\end{align}
Thus, the optimal solution to Problem \eqref{eq33} becomes
\begin{align}\label{p2}
\min\{\max\{R^{\min}, \bar{R}^\star\}, R^{\max}\}.
\end{align}

\emph{Complexity Analysis}: The computational complexity of solving Problem \eqref{eq33} is negligible.

\section{Numerical Results}
In our simulations, we make several key assumptions and parameter choices to set the conditions for investigation. We assume that the channel variances are $\sigma_{h}^2=1$, such that the link between SS and SD offers superior signal quality compared to the links between SD and the monitor, as well as between the monitor and SD, where $\sigma_{g}^2=\sigma_{f}^2=\sigma^2\leq1$. Additionally, we set $\sigma^2_{d}=\sigma^2_{m}=1$ for the additive Gaussian noises at both the SD and the monitor. The signal power at the SS is established as $p_s/\sigma^2=20$ dB, while the maximum transmission power of the jamming signal is capped at $p_m ^{\max}= 30$ dB. Our target outage probability at the SD is $\delta=0.05$, and we employ a parameter value of $W=5$, as referenced in \cite{KKWong21}. These assumptions and parameter selections collectively define the conditions for our simulation scenarios.

\begin{figure}[h]
\centering
\includegraphics[width=3.6in]{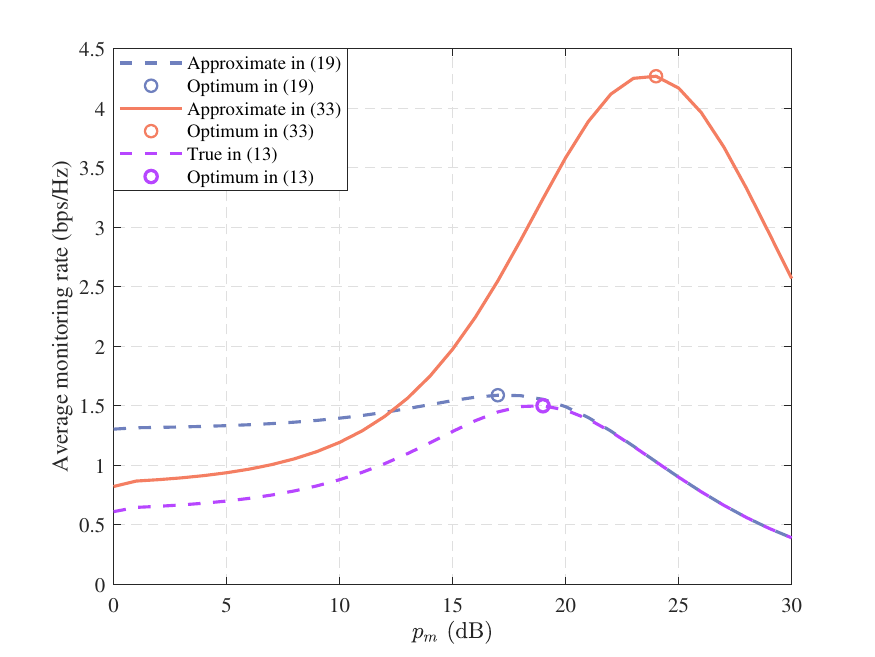}
\caption{True and approximate average monitoring rates versus $p_m$, where $N=8$ and $\sigma^2/\sigma^2_h=-18$ dB.}\label{fig:rVSp}
\end{figure}

In Fig.~\ref{fig:rVSp}, the results are provided for the true and approximate average monitoring rates across various values of $p_m$, considering a scenario with $N=8$ and $\sigma^2/\sigma^2_h=-18$ dB. In the legend, ``Approximate in \eqref{eq18}" corresponds to $\hat{R}_m$ in \eqref{eq18}, ``Approximate in \eqref{eq32}" represents $\tilde{R}_m$ in \eqref{eq32}, while ``True in \eqref{eq13}" represents $R_m$ in \eqref{eq13}. The results in Fig.~\ref{fig:rVSp} reveal that the optimal jamming powers for the monitor are $17$ dB, $24$ dB, and $19$ dB for ``Optimum in \eqref{eq18}", ``Optimum in \eqref{eq32}", and ``Optimum in \eqref{eq13}", respectively.

\begin{figure}[h]
\centering
\includegraphics[width=3.6in]{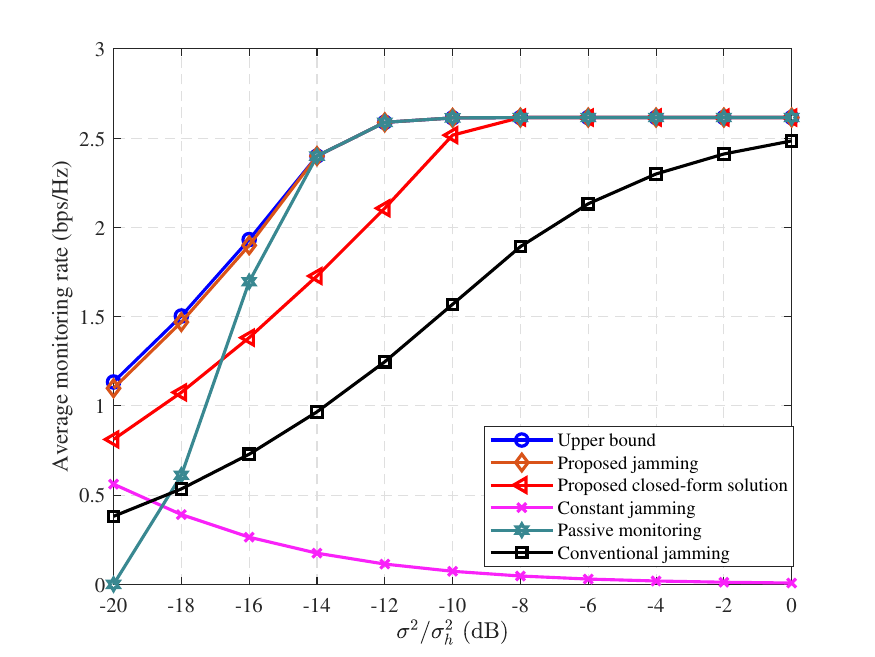}
\caption{Average monitoring rate versus $\sigma^2/\sigma^2_h$; performance comparison of different schemes, where $N=8$.}\label{fig:rVSs}
\end{figure}

In Fig.~\ref{fig:rVSs}, we  examine the impact of $\sigma^2/\sigma^2_h$ on the average monitoring rate. In the legend, ``Upper bound" means that the optimal solution of Problem \eqref{q11} is obtained through exhaustive search, while ``Proposed jamming" denotes the optimal solution of Problem \eqref{eq23}, ``Proposed closed-form solution" denotes the optimal solution of Problem \eqref{eq33}, ``Constant jamming" represents proactive monitoring with constant jamming power, where $p_m=P^{\max}_m=30$ dB, ``Passive monitoring" represents proactive monitoring without jamming.  Furthermore, ``Conventional jamming" represents the proactive monitoring with a conventional single antenna in \cite{XuJ16}.

Observing from Fig.~\ref{fig:rVSs}, we see that the  performance of ``Proposed jamming" scheme is nearly identical to that of the ``Upper bound" scheme. Additionally, it is evident that our proposed proactive monitoring schemes consistently outperform the  ``Conventional jamming" scheme. Notably, as depicted in Fig.~\ref{fig:rVSs}, when $\sigma^2/\sigma^2_h$ is high, the monitoring performance of our proposed schemes closely matches that of the ``Passive monitoring" scheme. This convergence occurs because, with a high $\sigma^2/\sigma^2_h$ ratio, the optimal jamming power is $p_m=0$.

\begin{figure}[h]
\centering
\includegraphics[width=3.6in]{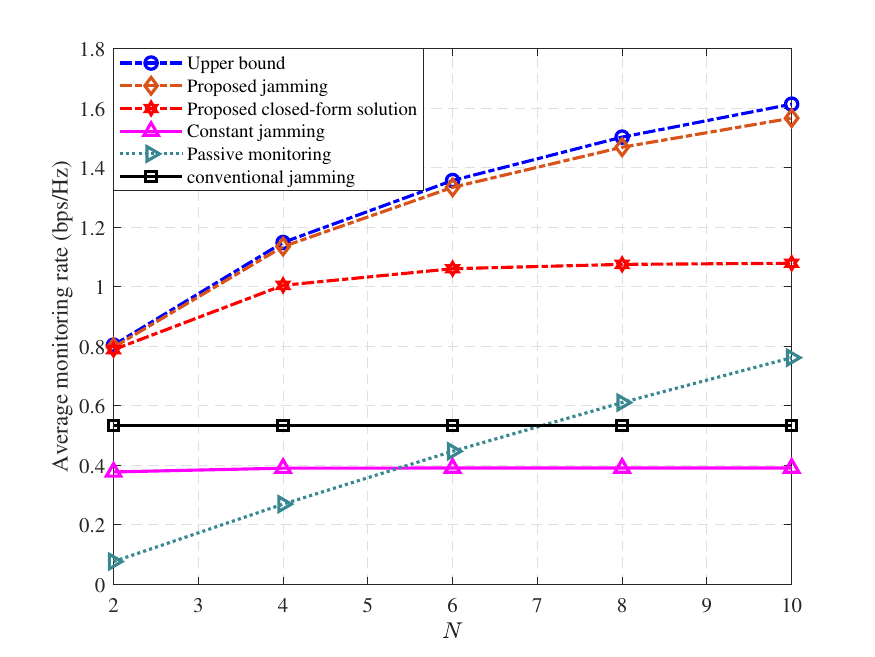}
\caption{Average monitoring rate versus $N$; performance comparison of different schemes, where $\sigma^2/\sigma^2_h=-18$ dB.}\label{fig:rVSn}
\end{figure}

Finally, in Fig.~\ref{fig:rVSn}, we demonstrate the impact of $N$ on the average monitoring rate with $\sigma^2/\sigma^2_h=-18$ dB. As seen from Fig.~\ref{fig:rVSn}, the monitoring efficacy of our proposed proactive monitoring schemes enhances as $N$ increases. This is due to the monitor's outage probability decreasing as $N$ goes up. It is noteworthy that the curve for the ``Constant jamming" scheme levels off when $N>4$, indicating little to no improvement in the outage probability as $N$ further increases.

\section{Conclusion}
In this paper, we focused on scenarios in which the monitor is equipped with a single fluid antenna for performance enhancement. We introduced two proactive monitoring schemes involving the design of jamming strategies. Our results consistently demonstrated that our proposed scheme surpasses other schemes, including proactive monitoring with a conventional single antenna, passive monitoring without jamming, and proactive monitoring with constant jamming power schemes.

\appendices
\section{Proof of Lemma 1}
From \eqref{eq26} and \eqref{eq27}, we know that $h(0)=1$ and $g(0)=0$, and $h(x)$ and $g(x)$ are both larger than 0 when $x>0$.

Taking the first-order partial derivative of $h(x)$ with respect to $x$, we can obtain
\begin{multline}\label{bq23}
\frac{\partial h(x)}{\partial x}=-\frac{1}{\ln2(1+x)^2}\left(1-\eta\left(1-e^{-\frac{x}{\Gamma(1-\mu^2)}}\right)^N\right)\\
-\frac{N}{\ln2(1+x)\Gamma(1-\mu^2)}\\
\times\left(\eta\left(1-e^{-\frac{x}{\Gamma(1-\mu^2)}}\right)^{N-1}e^{-\frac{x}{\Gamma(1-\mu^2)}}\right).
\end{multline}
We can readily know that $h(x)$ is a monotonically decreasing function with respect to $x$ when $x\geq 0$.

Similarly, taking the first-order partial derivative of $\log_2(1+x)g(x)$ with respect to $x$, we can obtain
\begin{multline}\label{bq23}
\frac{\partial g(x)}{\partial x}=\frac{N}{\ln2(1+x)\Gamma(1-\mu^2)}\\
\times\left(\eta\left(1-e^{-\frac{x}{\Gamma(1-\mu^2)}}\right)^{N-1}e^{-\frac{x}{\Gamma(1-\mu^2)}}\right)\\
+\log_2(1+x)\left(1-e^{-\frac{x}{\Gamma(1-\mu^2)}}\right)^{N-2}e^{-\frac{x}{\Gamma(1-\mu^2)}}\\
\times(N-2+e^{-\frac{x}{\Gamma(1-\mu^2)}}).
\end{multline}
We can readily know that $g(x)$ is a monotonically increasing function with respect to $x$ when $x\geq 0$ and $N\geq2$.
In our considered proactive monitoring system, we have $x\geq 0$ and $N\geq2$. As a result, we can conclude that $h(x)$ is a monotonically decreasing function with respect to $x$ and $g(x)$ is a monotonically increasing function.

Moreover, we have $h(0)=1$ and $g(0)=0$, and $h(+\infty)=0$ and $g(+\infty)=+\infty$. $h(x)-g(x)$ is larger than 0 over the interval $x\in[0,x^o)$, and smaller than 0 over the interval $x\in(x^o, +\infty]$, where $x^o=2^{R^o}-1$ and $R^o$ is the solution of $h(R^o)=g(R^o)$. Therefore, we can obtain that $F(x)$ is a monotonically increasing function over the interval $x\in[0,x^o]$, and monotonically decreasing function over the interval $x\in(x^o, +\infty]$, which concludes the proof.

\end{document}